\documentclass[12pt]{revtex4-1}
\usepackage{epsfig,epstopdf,amsmath,amssymb,graphicx,subfigure,verbatim,natbib}

\usepackage{color}

\usepackage{calligra}
\DeclareMathAlphabet{\mathcalligra}{T1}{calligra}{m}{n}
\DeclareFontShape{T1}{calligra}{m}{n}{<->s*[2.2]callig15}{}

\begin{document}

\author{John A. Brehm}
\email{john.a.brehm@vanderbilt.edu}
\affiliation{Department of Physics and Astronomy, Vanderbilt University, Nashville, Tennessee, 37235, United States}

\date{\today}

\begin{abstract}
The bulk photovoltaic effect (BPVE) is a phenomenon which creates a net electrical current from sunlight in a polar noncentrosymmetric material possessing a moderate band gap.  This effect is being explored as an alternative to traditional $pn$-junction solar cells to convert solar energy into electricity.  This paper assesses the possibility that hydrogenated Zintl-phase compounds consisting only of metal, metalloid, and hydrogen atoms, may be able to provide a BPVE response.  Towards this end, the magnitude of the BPVE shift current is calculated for each  of the compounds $AeTrTt$H ($Ae$ = Ca, Sr, and Ba; $Tr$ = Al and Ga; and $Tt$ = Si, Ge, and Sn). For this family of hydogenated Zintl compounds, maximum shift current responses are predicted that are up to eight times greater than that calculated for BiFeO$_\text{3}$, and have a significant response down to much lower photon energies than BiFeO$_\text{3}$ as well.  With band gaps below 1 eV, and with some members of this family of compounds exhibiting stability in air up to 770 K, some of these compounds may find use in photovoltaic devices.

\end{abstract}

\title{Predicted bulk photovoltaic effect in hydrogenated Zintl compounds} 
\maketitle


\section{Introduction.}

The bulk photovoltaic effect (BPVE) is a second-order nonlinear phenomenon in which electrons are excited from the valence band to the conduction band of a semiconducting or insulating compound by electromagnetic radiation, such as solar energy, with a net carrier velocity generated in particular crystallographic directions in real space, resulting in a net current.\cite{Sturman92p1,Young12p236601}    In contrast to traditional solar cell $pn$-junction materials physics, the BPVE does not require hetero-materials to create a current, nor is the photovoltage limited to the band gap of the material.   As such, it could allow for simpler manufacturing processes and produce materials with higher photovoltages.  The only materials requirements for a potential BPVE compound are that it be noncentrosymmetric and possess a band gap.  (However, in an environment of unpolarized sunlight, stricter requirements are that the compound be polar noncentrosymmetric and have a band gap at or below the visible range.)  The search for useful BPVE materials has led to predictions of the effect in traditional oxide perovskites (lead titanate and barium titanate),\cite{Young12p116601,Spanier16p611,Gu17p096601,Wang16p10419}  non-traditional oxide perovskites like BiFeO$_\text{3}$ and BiFeCrO$_\text{3}$,\cite{Young12p236601, Nechache11p202902,Kumar08p121915, Bhatnagar13p2835-1,Katiyar14p172904} hybrid organic-inorganic perovskites,\cite{Zheng15p31}  monoclinic non-perovskite chalcogenides,\cite{Brehm14p204704}  and proposed monochalcogenide chain arrangements.\cite{Rangel999p999,Cook17p14176}

\begin{figure}[h]
\vspace{30pt}
\includegraphics[scale = 0.65]{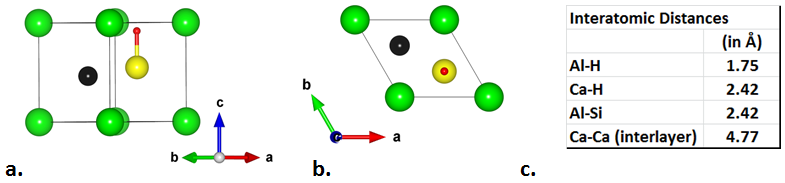}
\caption{\label{no1}The four-atom unit cell of CaAlSiH a.) side view and b.) top view.  (Green circles:  Ca; yellow circle:  Al; black circle: Si; and red circle: H.)  [These images are made with Vesta.\cite{VESTA13p1}]    c.)  Selected interatomic distances in CaAlSiH.  }
\end{figure}

In all the above referenced BPVE material examples, the ions in the compounds assume traditional cation-anion oxidation state roles.  Yet, materials chemistry is rich in compounds that depart from this paradigm. For example, Zintl phases consisting only of Group (I, II) metals and metalloids form well-known families  of compounds in which the more electronegative element(s) can be considered to be the anion(s).  In many instances, these anionic atoms interact and are termed a polyanion.  The $AeTrTt$ ($Ae$ = Ca, Sr, and Ba; $Tr$ = Al and Ga; and $Tt$ = Si, Ge, and Sn) compounds are such a family.    While all are metallic compounds, a hydrogenation process converts them into polar  semiconductors in the $P3m1$ space group with the chemical formula $AeTrTt$H and calculated  band gaps between 0.3-0.8 eV.\cite{Evans09p2068,Evans08p12139,Chihi13p1558,Lee08p195209,Kang12p153,Lu17p125117,Kranak09p1847,Bjoerling06p817, Bjoerling05p7269}   An optical absorption experiment on one of these compounds, SrAlSiH, has shown a band gap of 0.63 eV.\cite{Bjoerling05p7269}   CaAlSiH is representative of this family. Its unit cell, with the short Al-H bond length of 1.75~{\AA}, is shown in Figure \ref{no1}.     The bonding in this family of compounds has been described as either being the result of the formation of a formal valence state configuration of ($Ae$)$^\text{2+}$[$TrTt$H]$^\text{2-}$ or [$AeTrTt$]$^\text{1+}$(H$^\text{-}$).\cite{Lee08p195209,Lu17p125117,Kranak09p1847,Bjoerling05p7269}   The first model applies a strict adherence to the Zintl concept of a layered polyanion, which for this family of compounds is a combination of ``three-bonded [$Tr$-H]$^\text{-}$ and three-bonded lone electron pair $Tt$$^\text{-}$."\cite{Kranak09p1847}   On the other hand, the second model, as seen in a more recent study, has produced a band structure that seems to indicate that H orbital contributions mainly occur approximately 5 eV below the Fermi level, and mix only slightly with the $Tr$ and $Tt$ {$\pi$} bands nearer the Fermi level.\cite{Lu17p125117} Thus, this latter work regards [$AeTrTt$]$^\text{1+}$(H$^\text{-}$) as the proper designation.  In both cases, $Ae$, $Tr$, $Tt$, and H assume the formal oxidation states of 2$^\text{+}$, 3$^\text{+}$, 4$^\text{-}$, and 1$^\text{-}$ respectively.

In this work, as this $AeTrTt$H family of compounds meets the requirements for a BPVE material, the ``shift current," which has been demonstrated in earlier theoretical works to be the dominant mechanism for generating the BPVE in ferroelectrics, is calculated.\cite{Young12p236601,Young12p116601}  The Glass coefficient, which is used to assess the total shift current response in a material that is thick enough to absorb all penetrating light, is also calculated.\cite{Glass74p233}      One key reason for choosing this family of compounds to study is that many of its constituent elements are abundant in the crust of the earth and nontoxic, ($e.$ $g.$ Ca, Al, and Si in CaAlSiH).  Further, although these compounds have been produced in bulk by arc melting followed by hydrogenation, they could be amenable to thin film synthesis, since these elements do not present contamination or toxicity problems in reaction chambers (as other elements do, $e.$ $g.$  Li, S, Pb, and Se).  Also of importance to solar-cell device applications, members of the  $Ae$AlSiH sub-family of compounds have been reported to have decomposition temperatures in air  at or above 770 K.\cite{Lee08p195209}  Finally, all of the $AeTrTe$ family of compounds are low temperature superconductors.\cite{Evans09p064514}  As the $a$-lattice vector mismatch between each of these compounds and their corresponding hydrogentated phase is small ($e.$ $g.$ the $a$-lattice mismatch between CaAlSi and CaAlSiH is only 1.2 percent),  a seamless layered composite which has both BPVE and superconducting qualities might be achievable.

\section{Methods.}

A self-consistent field (SCF) density functional theory (DFT) calculation, within the generalized gradient approximation (GGA), is performed on each of the compounds $AeTrTt$H ($Ae$ = Ca, Sr, Ba), ($Tr$ = Al, Ga), and ($Tt$ = Si, Ge, Sn) using the Quantum Espresso computational package.\cite{Giannozzi09p395502} Furthermore, the same calculations are performed on a related compound, BaInGeH, for which it has been suggested that the H atom can have one of two locations in the unit cell and the In/Ge sites have mixed occupancy.\cite{Evans09p5602}  For consistency with the other compounds,  BaInGeH is treated as having only a single site for H, In, and Ge.  From the generated wavefunctions, band structure, orbital-projected density of states, shift current tensor elements, and Glass coefficient tensor elements are calculated. The equations for the latter two can be found in References \cite{Young12p116601}, \cite{Sipe00p5337}, and \cite{Baltz81p5590}.  The calculations for the shift current and Glass coefficient are performed using the code of Reference \cite{Young12p116601}. In the case of CaAlSiH, polarization is assessed using the Abinit computational package.\cite{Gonze02p478}    The atoms are modeled by norm-conserving pseudopotentials\cite{Rappe90p1227}  created with the Opium software package.\cite{opium17p1}   The norm-conserving pseudopotentials are designed with a 50 Ry plane-wave cutoff, and the SCF and polarization calculations use this cutoff value as well.   The atomic positions  are taken from References \cite{Evans09p2068}, \cite{Evans08p12139}, \cite{Lee08p195209}, \cite{Kranak09p1847}, and \cite{Bjoerling05p7269}, and the FIZ Karlsruhe ICSD database.\cite{Belsky02p364,FIZ13p1}

The family of compounds in this study has  $3m$ point group  symmetry.  The shift current response tensor is then represented in two dimensional matrix form as:\cite{Nye85p1}

\begin{eqnarray}
\sigma  = \begin{bmatrix}
       0 & 0 & 0 & 0 & \sigma_{xzX}  & \sigma_{xyX} \\ 
       \sigma_{xxY} & \sigma_{yyY} & 0 & \sigma_{yzY} & 0 & 0            \\[0.3em]
       \sigma_{xxZ} & \sigma_{yyZ} & \sigma_{zzZ} & 0 & 0  & 0\\[0.3em]
     \end{bmatrix}
\end{eqnarray}

where where $\sigma_{xxZ}$ = $\sigma_{yyZ}$, and -$\sigma_{xxY}$ = $\sigma_{yyY}$.   The off diagonal terms, ($\sigma_{xzX}$, $\sigma_{xyX}$ and $\sigma_{yzY}$), give canceling contributions in the presence of unpolarized light.\cite{Brehm14p204704}

\section{Results and discussion.}

\begin{figure}[h]
\vspace{30pt}
\includegraphics[scale = 0.5,]{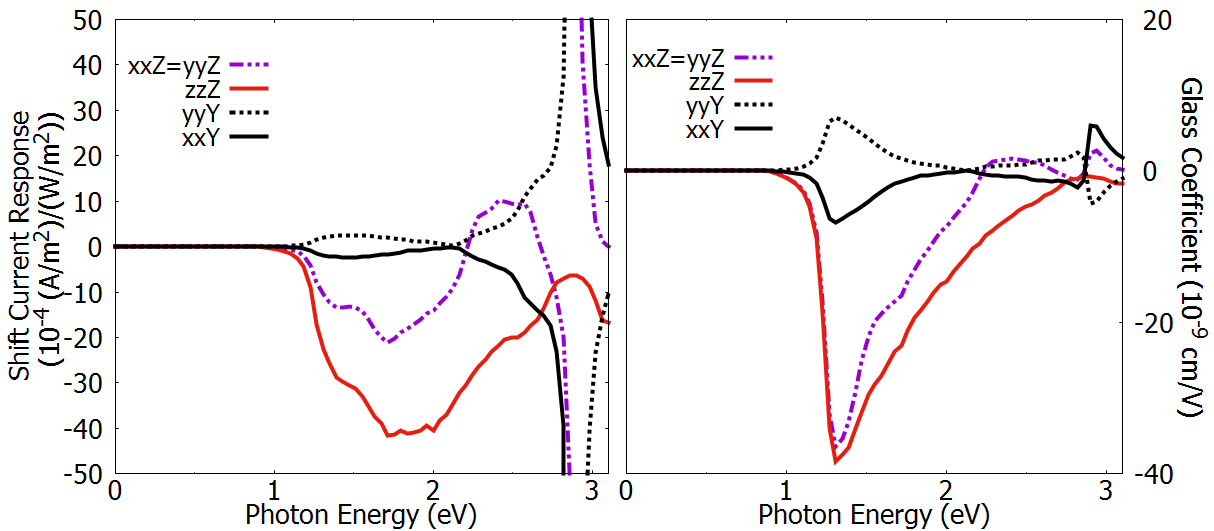} \hspace*{1cm}
\caption{\label{no2}Shift current and Glass coefficient responses for CaAlSiH as a function of electromagnetic radiation.  The legend entries are interpreted as follows:  $zzZ$ means polarized light from the $zz$ direction inducing a current in the $Z$ Cartesian direction.}
\end{figure}

Table I presents the calculated maximum shift current and maximum Glass responses, as well as the calculated GGA band gap.  It is clearly seen that all of these compounds produce a significant calculated shift current and glass coefficient response.  Except for those compounds with $Tt$ = Sn, both the shift current and Glass coefficient responses are 5-8 times greater than comparable values calculated for the benchmark compound BiFeO$_\text{3}$.\cite{Young12p236601}   Figure \ref{no2} shows these responses for CaAlSiH.    The responses are significant between 1 and 3 eV.  Inspection of the response shows that -$\sigma_{xxY}$ = $\sigma_{yyY}$  and that  $\sigma_{xxZ}$ = $\sigma_{yyZ}$, respecting the crystal symmetry.  In addition, the  $\sigma_{zzZ}$ response is approximately twice that of $\sigma_{xxZ}$ = $\sigma_{yyZ}$.   These behaviors are representative of most of the $AeTrTt$H compounds. 

\begin{table*} 
\caption{\label{thedata} Calculated GGA band gaps, maximum shift current response, and maximum Glass coefficient for $AeTrTt$H compounds.  The last block of compounds is for presented for comparison.  }

\begin{tabular}{|c|c|c|c|} 
\hline
\hline
                 & Calculated & Max. shift & Max. Glass \\ 
                 & GGA Band Gap & current density & coefficient \\ 
Compound &  (eV) &  ($\times$10$^\text{-4}$ (A/m$^\text{2}$)/(W/m$^\text{2}$)) & ($\times$10$^\text{-9}$ cm/V) \\  \hline

BaAlGeH & 0.81 & -36 & -24 \\
BaAlSiH & 0.69 & -40 & -25 \\
BaGaGeH & 0.42 & -33 & -30 \\
BaGaSiH & 0.53 & -36 & -30 \\
BaGaSnH & 0.28 & -22 & -20 \\
BaInGeH & 0.63 & -39 & -10 \\  \hline
CaAlSiH & 0.35 & -42 & -38 \\
CaGaGeH & 0.58 & -34 & -22 \\
CaGaSiH & 0.28 & -37 & -39 \\
CaGaSnH & 0.33 & -14 & -5 \\ \hline
SrAlGeH & 0.74 & -33 & -29 \\
SrAlSiH & 0.62 & -37 & -32 \\
SrGaGeH & 0.68 & -36 & -25 \\
SrGaSiH & 0.47 & -37 & -35 \\
SrGaSnH & 0.59 & -19 & -7 \\  \hline
BiFeO$_\text{3}$  \cite{Young12p236601} & 2.5 & 5 & 5 \\
KBNNO  \cite{Wang15p165214} & 1.30 (HSE) & 25 & 5 \\
LiAsSe$_\text{2}$   \cite{Brehm14p204704} & 0.77 & -98 & -42 \\
LiAsS$_\text{2}$ \cite{Brehm14p204704} & 1.07 & -49 & -21 \\
\hline
\hline
\end{tabular}
\end{table*}

\begin{figure}[h]
\vspace{20pt}
\includegraphics[scale = 0.55,]{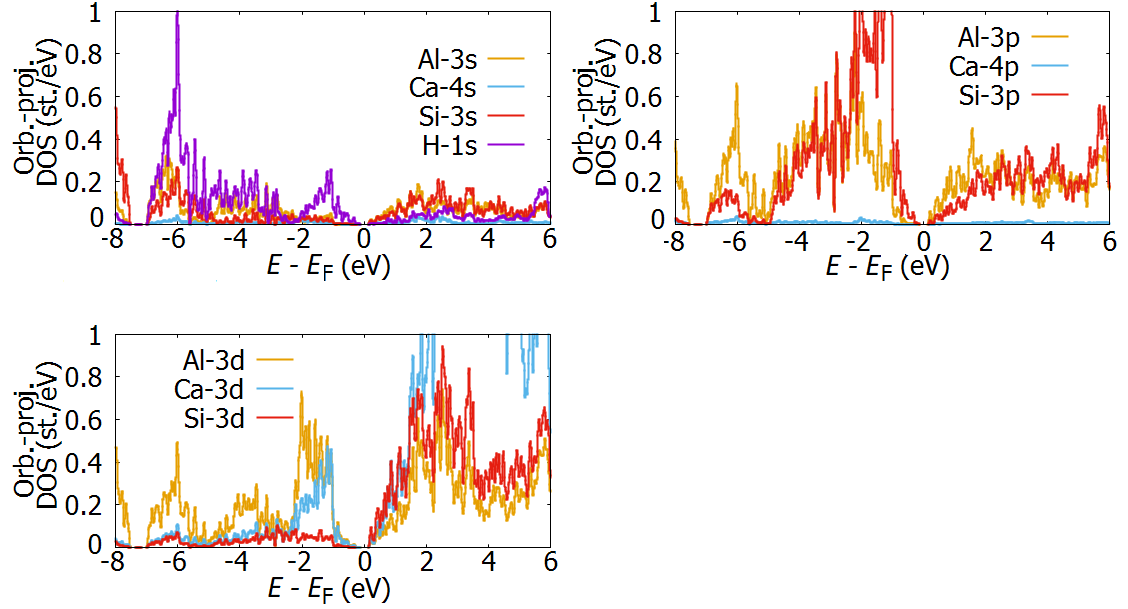} \hspace*{1cm}
\caption{\label{fig3}Orbital-projected density of states plots for CaAlSiH for $s$-orbitals,  $p$-orbitals, and $d$-orbitals.}
\end{figure}

The orbital-projected density of states is calculated for CaAlSiH and shown in Figure \ref{fig3}. It is noted that the elements Ca, Al, and Si have been modelled with pseudopotentials that treat the nominally empty 3$d$-orbitals accurately.  The plots show that the valence band edge is composed primarily of H-1$s$ and Si-3$p$ states.  This agrees with previous work in published in Reference \cite{Lee08p195209}.  As well, at the conduction band edge, the Al-3$p$ states are more prominent in the DOS than the other $s$- and $p$- states, again in agreement with \cite{Lee08p195209}.  The electronic structure presented here differs from previous reports in that that the $d$-states of Ca, Al, and especially Si are present at the conduction band edge and, individually, are comparable in magnitude, if not greater in magnitude, than the Al-3$p$ states.  This is of importance since these conduction band 3$d$ states could play an important role in the shift current and Glass coefficient responses of CaAlSiH.   

The plots in Figure \ref{fig3} also show strong orbital interaction between H-1$s$ states and Al-3$s$, -3$p$, and -3$d$ states between -7 and -5 eV.  To a lesser extent, the  Si 3$s$ and 3$p$ orbitals interact with H-1$s$ in this energy range.  However, H-1$s$  has a significant density between -2 and 0 eV as well, interacting with the 3$d$ orbitals of Ca and Al and the 3$p$ orbitals of Al and Si.  The combination of interactions in these two regions leads to a more complicated designation of a formal valence state configuration of $AeTrTt$H as being  either ($Ae$)$^\text{2+}$[$TrTt$H]$^\text{2-}$ or [$AeTrTt$]$^\text{1+}$(H$^\text{-}$).

\begin{figure}[h]
\vspace{10pt}
\includegraphics[scale = 0.65,]{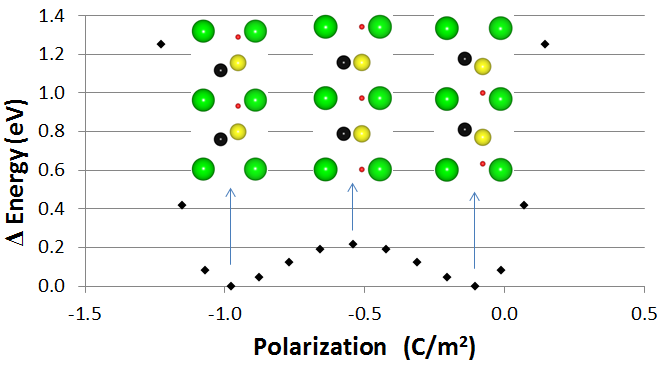} \hspace*{1cm}
\caption{\label{rot}Polarization versus SCF energy plot.  The $xz$-plane depictions of the double unit cells show the atoms at the ground state configurations (left and right), and at a $z$-coordinate centrosymmetric configuration (middle).  All energy differences are relative to the ground state. }
\end{figure}

Polarization calculations for CaAlSiH show that the sum of the total Berry phase and ionic polarization is negligible in the $x$- and $y$-directions (less than 6 $\times$ 10$^\text{-4}$ C/m$^\text{2}$), but is significant in the $z$-direction:  the magnitude of the polarization is P = P$_\text{$z$}$ = 0.11 C/m$^\text{2}$. By comparison, for BaTiO$_\text{3}$, P = 0.90  C/m$^\text{2}$.\cite{Wang03p1719}  This result is further evidence that the sizes of the shift current and Glass coefficient responses do not vary proportionally to the magnitude of polarization.\cite{Brehm14p204704}  In order to determine if the polarization could be easily flipped in direction, (and thereby the material may be  considered a ferroelectric), two series of SCF calculations were performed on the unit cell constructed so that it had $z$-direction mirror symmetry.   The reduced cell coordinates for these calculations were  (0, 0, 0), (1/3 , 2/3, 1/2), (2/3, 1/3, 1/2) for Ba, Al, and Si respectively.  The H atom was either placed in the same $xy$-plane as Al (and Si) or  Ba.   For the case when the H atom was placed in the same $xy$-plane as the Al and Si atoms, three calcualtions were performed with H having the reduced coordinates (1/2, 1/2, 1/2), (1/3, 1/3, 1/2),  and (0, 0, ~1/2). The difference in energies between these configurations and  the ground state configuration is termed the well depth and is calcualted to be 6.23 eV, 3.31 eV, and 1.58 eV respectively.  In comparison, the well depths calculated for the ferroelectrics PbTiO$_\text{3}$ and LiNbO$_\text{3}$ have been reported at 0.16-0.22 eV  and at 0.25 eV respectively.\cite{Wu04p104112,Ye16p134303}  Thus, it  appears unlikey that CaAlSiH would be classified as a ferroelectric in the traditional sense.   However, two calculations were performed on unit cells in which the H atom was placed in the same $xy$-plane as Ba:  the first at  (1/2, 1/2, 0), and the second at (1/3, 2/3, 0).   The   difference in energy between these configurations and  the ground state configuration was 1.14 eV higher for the former, but only 0.21 eV higher for the latter.  This is much more in line with the other two ferroelectrics listed above.  It seems likely that the H atom would likely pass through the layer of Ba atoms to reverse polarization under an electric field than pass through an Al-Si layer.  There might be more complicated ionic movements of the Al and Si atoms in which all three coordinates change for both and which result in a smaller well depth energy, but these have not been found in the current study.   Figure \ref{rot} shows the double well polarization versus energy plot for CaAlSiH when the H atom movement is considered to be towards the plane of Ba atoms.  Note, the figure is plotted in consideration of the modern theory of polarization in which the quantum of polarization,  P$_\text{$q$}$, is taken into account.\cite{Resta07p31}  For this compound, P$_\text{$q$}$ = 1.08 C/m$^\text{2}$ in the $z$ direction.  Further,  the $z$-symmetric unit cell has a P=  P$_\text{$z$}$ = -0.54 C/m$^\text{2}$.   This is indicative of a P$_\text{$q$}$/2 + $n$ P$_\text{$q$}$  system.\cite{Spaldin12p2,Resta07p31}

\section{Conclusions.}

The calculations performed in this work predict that the $AeTrTt$H ($Ae$ = Ca, Sr, and Ba; $Tr$ = Al and Ga; and $Tt$ = Si, Ge, Sn)  and BaInGeH hydrogenated Zintl compounds will produce significant calculated second order opto-electronic responses when subjected to visible light.  The calculated shift current and Glass coefficient responses are up to eight times greater than those calculated for BiFeO$_\text{3}$, and the strong PV response spans  the visible spectrum.  It is remarkable  that none of the elements is a traditional anion from the Groups 14, 15 or 16.  The shift current and Glass responses were highlighted for CaAlSiH, not only because they represent the highest or near-highest responses for this family of compounds, but because all of the elements are plentiful and non-toxic.  Lastly, the importance of treating empty orbitals accurately in the calculations is noted, not only for BPVE assessment, but for basic electronic structure as well.

\section{Acknowledgements.} 
This work was supported by the U.S. Department of Energy grant DE-FG02-09ER46554.  JAB would like to thank Dr. Ulrich H{\"a}ussermann of the Stockholm University's Department of Materials and Environmental Chemistry for instructive conversations regarding the $AeTrTt$H family of compounds.  JAB would like to thank Dr. Andrew M. Rappe of the University of Pennsylvania's Department of Chemistry for reviewing this manuscript, and for providing the computational code and computer resources necessary to perform the calculations.   Computational support was provided by the NERSC Center of the U.S. DOE.   JAB would like to thank Dr. Steve Young  at the Sandia National Lab in Livermore, Ca. and Dr. Fan Zheng at the Lawrence Livermore National Lab in Berkeley, Ca. for useful conversations concerning the input, output, and compiling of the shift current code.

\end{document}